\documentclass[twocolumn,superscriptaddress]{revtex4-2}

\usepackage{amsmath}
\usepackage{graphicx}
\usepackage{xcolor}

\usepackage{lineno,hyperref}
\modulolinenumbers[5]

\graphicspath{{./fig/}}

\begin{document}


\title{Learned discretizations for passive scalar advection in a 2-D turbulent flow}

\author{Jiawei Zhuang}
\affiliation{School of Engineering and Applied Sciences, Harvard University, Cambridge, MA}
\affiliation{Google Research, 1600 Amphitheatre Pkwy, Mountain View, CA}
\author{Dmitrii Kochkov}
\affiliation{Google Research, 1600 Amphitheatre Pkwy, Mountain View, CA}
\author{Yohai Bar-Sinai}
\affiliation{School of Engineering and Applied Sciences, Harvard University, Cambridge, MA}
\affiliation{Google Research, 1600 Amphitheatre Pkwy, Mountain View, CA}
\author{Michael P. Brenner}
\affiliation{School of Engineering and Applied Sciences, Harvard University, Cambridge, MA}
\affiliation{Google Research, 1600 Amphitheatre Pkwy, Mountain View, CA}
\author{Stephan Hoyer}
\affiliation{Google Research, 1600 Amphitheatre Pkwy, Mountain View, CA}


\begin{abstract}
The computational cost of fluid simulations increases rapidly with grid resolution. This has given a hard limit on the ability of simulations to accurately resolve small scale features of complex flows. Here we use a machine learning approach to learn a numerical discretization that retains high accuracy even when the solution is under-resolved with classical methods. We apply this approach to passive scalar advection in a two-dimensional turbulent flow. The method maintains the same accuracy as traditional high-order flux-limited advection solvers, while using $4\times$ lower grid resolution in each dimension. The machine learning component is tightly integrated with traditional finite-volume schemes and can be trained via an end-to-end differentiable programming framework. The solver can achieve near-peak hardware utilization on CPUs and accelerators via convolutional filters. Code is available at \url{https://github.com/google-research/data-driven-pdes}.
\end{abstract}

\maketitle



\section{Introduction}

A key problem in the numerical simulation of complex phenomena is the need to accurately resolve spatiotemporal features over a wide range of length scales. For example, the computational requirement for simulating a high Reynolds number fluid flow scales like ${\rm Re}^3$, implying that a tenfold increase in Reynolds number requires a thousand fold increase in computing power. Over the past decades, the extra computing power made available through Moore's law has been used to increase grid resolution dramatically, leading to breakthroughs in turbulence modeling \cite{jimenez2019}, weather prediction \cite{bauer2015}, and climate projection \cite{schneider2017a}. Nonetheless, there is still a formidable gap towards resolving the finest spatial scales of interest \cite{neumann2019}, especially with the recent slow-down of Moore's Law \cite{theis2017,shalf2020}. Machine learning has given a potential way out of this conundrum, by training low-resolution models to learn the rules from their high-resolution counterparts \cite{schneider2017b,reichstein2019,kutz2017,duraisamy2019}. The learned models aim to produce high-fidelity simulations using much less computational resources. Incorporating machine learning into numerical models also facilitates the adoption of emerging hardware, considering that the fastest growth in computing power now relies on domain-specific architectures like Graphical Processing Units (GPUs) \cite{hatfield2019} and Tensor Processing Units (TPUs) \cite{hennessy2019,dean2019} that are optimized for machine learning tasks.

Recently we introduced {\sl data driven discretizations} \cite{bar2019} to learn numerical methods that achieve the same accuracy as traditional finite difference methods but with much coarser grid resolution. These methods are equation specific, and require training a coarse resolution solver with high resolution ground truth simulations. Since the dynamics of a partial differential equation is entirely local, the high resolution simulations can be carried out on a small domain. We demonstrated the method with a set of canonical one-dimensional equations, demonstrating a $4 \sim 8 \times$ upscaling of effective resolution \cite{bar2019}. Here we extend this methodology to two-dimensional advection of passive scalars in a turbulent flow, a canonical problem in physics \cite{shraiman2000} and a classic challenge in atmospheric modeling \cite{rastigejev2010}. We show that machine-learned advection solver can use a grid with $4\times$ coarser resolution than classic high-order solvers while still maintaining the same accuracy.

\section{Data-driven solution to advection equation}

\subsection{Advection equation} 

We consider the advection of a scalar concentration field $C(\vec{x}, t)$ under a specified velocity  field $\vec{u}(\vec{x},t)$:
\begin{equation}
\frac{\partial C }{\partial t} + \nabla \cdot (\vec{u} C)  = 0
\label{eq:advection}
\end{equation}
If the prescribed velocity field is divergence-free
\begin{equation}
\nabla \cdot \vec{u} = 0,
\end{equation}
then, Eq. (\ref{eq:advection}) reduces \cite{leveque1996}
\begin{equation}
\frac{\partial C }{\partial t} + \vec{u} \cdot \nabla C = 0.
\end{equation}

A classical Eulerian scheme uses discretizations of the spatial derivative $\frac{\partial C}{\partial x}$, often in a form of:
\begin{equation}
\left.\frac{\partial C}{\partial x}\right|_{x=x_i} = \sum_{j=-k}^{k} \alpha_j C_{i+j}
\label{eq:fd-coefficients}
\end{equation}
where $\{x_1, .., x_N\}$ is the spatial grid points, $C_j$ is the concentration at point $x_j$, and $\{\alpha_{-k},...,\alpha_{k}\}$ are predefined finite-difference coefficients. For example, a first-order forward difference $\frac{C_{i+1} - C_i}{\Delta x}$ (where $C_{i+1}$ is in the upwinding direction) leads to the upwind scheme. Sophisticated high-order methods with flux limiters will choose different coefficients depending on local fields \cite{sweby1984}. Extension to two-dimensions can be done by either operator splitting (solve for each dimension separately) \cite{lin1996} or a true two-dimensional discretization \cite{ullrich2010}.

Although high-order Eulerian schemes are highly accurate under idealized flows \cite{colella2008}, their accuracy breaks down to first-order under turbulent or strongly sheared flows, resulting in significant numerical diffusion \cite{rastigejev2010}. Adaptive mesh refinement can reduce such numerical diffusion \cite{semakin2016}, but increases software complexity. Lagrangian methods avoid numerical diffusion \cite{stohl2005}, but have inhomogeneous spatial coverage and also difficulties in dealing with nonlinear chemical reaction \cite{eastham2017}. Semi-Lagrangian approaches involve remapping from a distorted Lagrangian mesh to a regular Eulerian mesh \cite{lauritzen2010}, and such remapping step exhibits similar numerical diffusion as Eulerian methods. Flow-map approaches \cite{kulkarni2019} can achieve Lagrangian-like accuracy on a Eulerian mesh, but need to solve for the advection trajectory over multiple steps and requires a special treatment to incorporate additional terms (e.g. chemical reaction) between advection steps. Different from existing methods, here we aim to develop an ultra-accurate advection solver under the requirements of: (1) a strictly Eulerian framework on a fixed grid, (2) explicit time-stepping, and (3) only relying on the current state to predict the next time step.

\subsection{Learning optimal coefficients} 
\label{section:learning-coef}

Instead of using predefined rules to compute finite-difference coefficients (Eq. \ref{eq:fd-coefficients}), our {\sl data driven discretizations} \cite{bar2019} predict the local-field-dependent coefficients $\vec{\alpha}=\{\alpha_{-k},...,\alpha_{k}\}$ via a convolutional neural network:
\begin{equation}
\vec{\alpha} = f(C, \vec{u}; W)
\end{equation}

The coefficients $\vec{\alpha}_{|x=x_j}$ depend on the local environment around $x_j$, with the inputs to the neural network being the neighboring fields $\{C_j, C_{j\pm1},...\}$ and $\{\vec{u}_j, \vec{u}_{j\pm1},...\}$. For simplicity of presentation, here we use 1-D indices $\{j, j\pm1, ...\}$ to denote spatially adjacent points. For 2-D advection problems, this computation involves 2-D convolution across both $x$ and $y$ dimensions. We learn the neural network weights $W$ by minimizing the difference between the machine learning prediction and the true solution. 

\begin{figure*}
  \centering
  \includegraphics[width=0.8\linewidth]{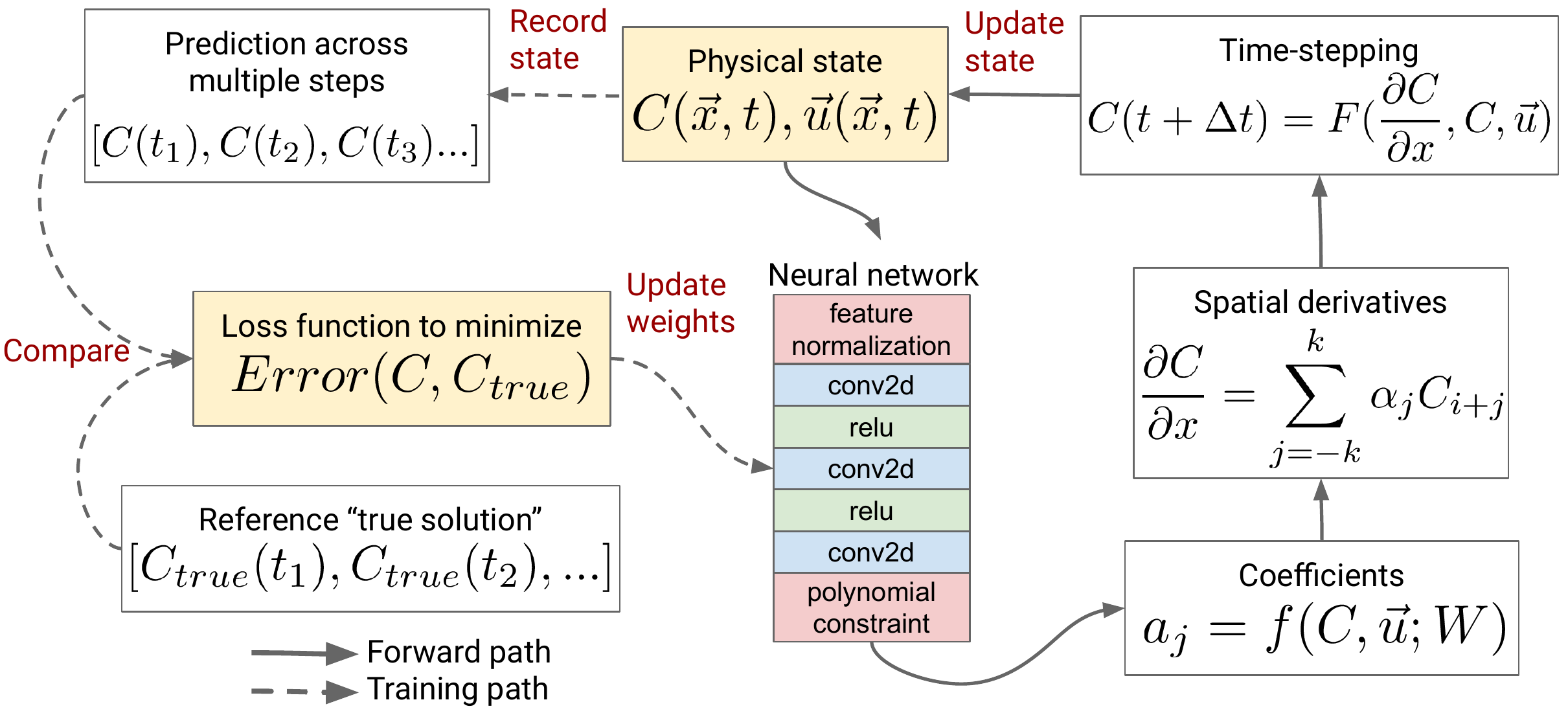}
  \caption{
  \textbf{End-to-end learning framework with differential programming.} During training, the model is optimized to predict future concentrations across multiple time steps, based on a precomputed dataset of snapshots from high resolution simulations. During inference, the optimized model is repeatedly applied to predict time evolution. The neural network component contains a stack of 2-D convolutional layers with ReLU activation functions (degraded to 1-D convolution for 1-D problems.). Physical constraints are imposed before and after the convolutional layers (Section \ref{section:constraint}). In the ``Time-stepping" block, $H$ is the advection operator that computes the concentration update based on the machine-learning estimate of spatial derivatives.
  }
  \label{fig:learning-framework}
\end{figure*}

Fig.~\ref{fig:learning-framework} shows the forward solver workflow and training framework. During the forward solve, we replace the computation of finite-difference coefficients with a convolution neural network, while still using classic approaches for the rest of the steps (computing the advection flux and doing the time-stepping). During training, we accumulate the forward solver prediction results over 10 time steps and then compare to the reference solution over this time period, by computing the mean absolute error (MAE) over the entire spatial domain between the two time series: 
\begin{equation}
{MAE} = \frac{1}{N \cdot M}\sum_{i=1}^{N}\sum_{j=1}^{M}\left| C^{predict}_j(t_i) - C^{true}_j(t_i) \right|
\end{equation}

The MAE is used as the loss function for neural network training \footnote{Using this MAE loss indicates that we require point-wise agreement between the machine learning prediction and the reference true solution, on every time step. This criteria is relevant for atmospheric transport modeling where the goal is to simulate the instantaneous, deterministic concentration field \cite{rastigejev2010}. For turbulence modeling, matching the high-order statistics might be more desirable than requiring point-wise agreement of instantaneous fields. This would require changes in the loss function.}. We find that using this multi-step loss function (as opposed to a single time step) stabilizes the forward integration, similar to the findings by \cite{brenowitz2018}. In our experiments, we found using MAE resulted in slightly more accurate predictions than using mean square error (MSE), but the difference was not large.

The training of a neural network inside a classic numerical solver is made possible by writing the entire program in a differentiable programming framework \cite{innes2019}, which allows efficient gradient-based optimization of arbitrary parameters in the code using automatic differentiation (AD) \cite{baydin2017}. AD tools have a long history, dating back to Fortran 77 \cite{bischof1996}. Recent developments of AD frameworks, such as TensorFlow \cite{abadi2016}, PyTorch \cite{paszke2017}, JAX \cite{jax}, Flux.jl \cite{innes2018}, and Swift \cite{swift}, are even easier to program and support hardware accelerators like GPUs and TPUs. Those developments make it easier to incorporate machine learning into scientific computing code (e.g.\ \cite{schoenholz2019}). We implemented our advection solver in TensorFlow Eager \cite{agrawal2019}.

\subsection{Baseline solver and reference solution}

As a baseline method, we use the second-order VanLeer advection scheme with a monotonic flux limiter \cite{lin1994}. To obtain the reference ``true" solution, we run the baseline advection solver at sufficiently high resolution to ensure the solution has converged. We then down-sample the high-resolution results using conservative averaging, to produce the training and test datasets for our machine-learning-based model on a coarse grid. 

We remark that although higher-order schemes with more advanced limiters would be more accurate, any flux-limited high-order schemes break to first-order under turbulent flows in order to ensure monotonicity \cite{rastigejev2010}. Starting from second-order, increasing the spatial resolution is generally more effective than further improving the solver order or the limiter \cite{smaoui2001,hassanzadeh2009}.

\subsection{Physical constraints}
\label{section:constraint}

There is growing emphasis on embedding physical constraints into the design of machine learning methods. This is typically done either either by adding ``soft'' constraints as terms the loss function~\cite{raissi2018,raissi2019}, or ``hard'' constraints in the model architecture~\cite{Bezenac2019-ia,bar2019,Um2020-zf,Pathak2020-mz,Frezat2020-zv,Ling2016-qk}.
Since here we only replace a small component in the numerical solver with machine learning, we can impose arbitrary physical constraints before and after the neural network components.
Using hard constraints allows the machine learning algorithm to focus on approximation problems, by imposing physical consistency requirements by construction. In particular, we require:

(1) {\sl Finite-volume representation for mass conservation}. We compute the flux across grid cell boundaries, and then apply the flux to update the concentration fields $C_{i}$. This ensures that mass is exactly conserved. The machine-learning estimate of spatial derivatives $\frac{\partial C}{\partial x}$ is used for obtaining the optimal interpolation values $C_{i+\frac{1}{2}}$ at cell boundaries, which is then used for calculating the flux via $u_{i+\frac{1}{2}}C_{i+\frac{1}{2}}$.

(2) {\sl Polynomial accuracy constraints}. Following \cite{bar2019}, we can force the machine-learning-predicted coefficients to satisfy an $m$-th order polynomial constraint, so that the approximation error decays as $O(\Delta x^m)$. This ensures that if the learned discretization is fit to solutions that are smooth on the scale of the mesh, we will recover classical finite-difference methods. In our experiments, we find that a first-order constraint gives the best result on coarse grids. This preserves a balance between accuracy constraints and model flexibility that may be particular valuable in non-monotonic regions, where higher order advection schemes often revert to first-order upwinding~\cite{sweby1984}.
First-order accuracy requires $\sum_{j=-k}^{k}\alpha_j=0$, and can be enforced by applying an affine transformation to the original neural network output (our implementation), or by having the neural network only output $\{\alpha_{-k}, ..., \alpha_{k-1}\}$ and solving for the last $\alpha_{k}$.
We choose the constant vector in the affine transformation to match a centered, first order scheme (equal weight on the two nearest grid cells).
Accordingly, our randomly initialized neural net at the start of training produces interpolation coefficients that are very close to a centered, first order scheme.

(3) {\sl Input feature normalization}. Before feeding the current concentration field $C$ to the neural network, we normalize it globally to $[0,1]$. This ensures that the overall magnitude of the concentration does not affect the prediction of finite-difference coefficients, and thus our solver satisfies the ``semi-linear" requirement for advection schemes that $H(aC+b)=aH(C)+b$ where $H$ is the advection operator and $\{a, b\}$ are constants (Eq 2.12-2.13 of \cite{lin1996}). Without such normalization, we find that the trained model diverges quickly during the forward integration.

\subsection{Other choices of learned terms}

Our training framework can be easily adapted to learn other parameters besides the finite-difference coefficients. In this section, we describe other approaches that we experimented with but did not choose.

Numerical methods introduce artificial numerical dissipation, so it is natural to consider adding explicit corrections to diffusion.
One of the earliest flux-correct transport (FCT) algorithms \cite{boris1973} includes an anti-diffusion coefficient of 1/8 as a correction term, though the choice of 1/8 was subjective and it was later acknowledged that such correction should better be velocity- and wavenumber- dependent~\cite{book2012}.
We considered learning diffusive correction directly, in the form:
\begin{equation}
\frac{\partial C }{\partial t} + \nabla \cdot (\vec{u} C) + \left(
D_{xx}\frac{\partial^2 C}{\partial x^2} +
D_{xy}\frac{\partial^2 C}{\partial x \partial y} + 
D_{yy}\frac{\partial^2 C}{\partial y^2}
\right) = 0,
\end{equation}
where the (anti-)diffusion coefficients $\vec{D} = \{D_{xx}, D_{xy}, D_{yy}\}$ are computed by a convolutional neural network $\vec{D} = f(C, \vec{u}; W)$, while the advection-diffusion equation itself is still solved by a traditional high-order finite volume method.
The idea resembles learning the Reynolds stress tensor~\cite{duraisamy2019} in a Reynolds averaged Navier Stokes (RANS) simulation. As in Section \ref{section:learning-coef}, here the neural network is trained by minimizing the difference between the model prediction and the reference solution. In practice, we found that this learned diffusion model achieves about $3 \times$ upscaling compared to the second-order baseline solver, but performs slightly worse than our original approach of learning finite-difference coefficients (Section \ref{section:learning-coef}) that can achieve $4 \times$ upscaling.

We also experimented with other learned terms, including (1) a pure machine learning approach, by having the neural network directly predict the concentration at the next step $C(t + \Delta t)$ based on the current state $C(t)$ and $\vec{u}(t)$; and (2) having the neural network directly predict the spatial derivative $\frac{\partial C}{\partial x}$ instead of the finite-difference coefficients $\vec{\alpha}$ that need to be further multiplied with the concentration field $C$ to obtain the spatial derivative. We found those methods to be unstable due to the lack of physical constraints (Section \ref{section:constraint}).

\section{Numerical results}

We apply the {\sl data driven discretization} to one- or two- dimensional advection. Two-dimensional advection is highly relevant for atmospheric modeling, as the vertical dimension can be decoupled from the horizontal dimensions and solved independently \cite{lin1996}.

The performance of our learned advection solver (the ``neural network model" hereafter) depends on the hyperparameters of the convolutional neural network component. For simplicity, this section only presents the results with the default hyperparameter configuration. For 1-D problems, we use 4 convolutional layers and 32 filters in each layer; For 2-D problems, we use 10 convolutional layers and 128 filters in each layer. All cases use a 3-point finite difference stencil ($k=1$ in Eq. \ref{eq:fd-coefficients}). The impact of hyperparameters on model accuracy and computational speed is further examined in Section \ref{section:speed}. We use the Adam optimizer \cite{kingma2014} with default parameters for neural network training. Our simple convolutional neural network achitecture already achieves a high accuracy, without additional operations like residual connections and batch normalization. 

\subsection{1-D advection under constant velocity}

We first show that our neural network model can achieve near-perfect result for a canonical test problem: 1-D advection constant velocity \cite{lin1994}. We consider a periodic 1-D grid of 32 grid points. The concentration field is shifted by a constant distance per time step, determined by the Courant–Friedrichs–Lewy (CFL) number $\frac{u\Delta t}{\Delta x}$. We set CFL = 0.5 ($\Delta x = 1$, $\Delta t = 0.5$, $u = 1$), so that the concentration field is shifted by half grid box every time step, and returns to the original position after every 64 time steps.

To generate training data, we initialize 30 square waves with heights randomly-sampled from $[0.1, 0.9]$ and widths from $2 \sim 8$ grid points. Test data are randomly sampled from the same range of width and height. The reference ``true" solution is generated by the baseline solver at $8\times$ resolution (256 grid points) and down-sampled to the original coarse grid.

\begin{figure}
  \centering
  \includegraphics[width=\linewidth]{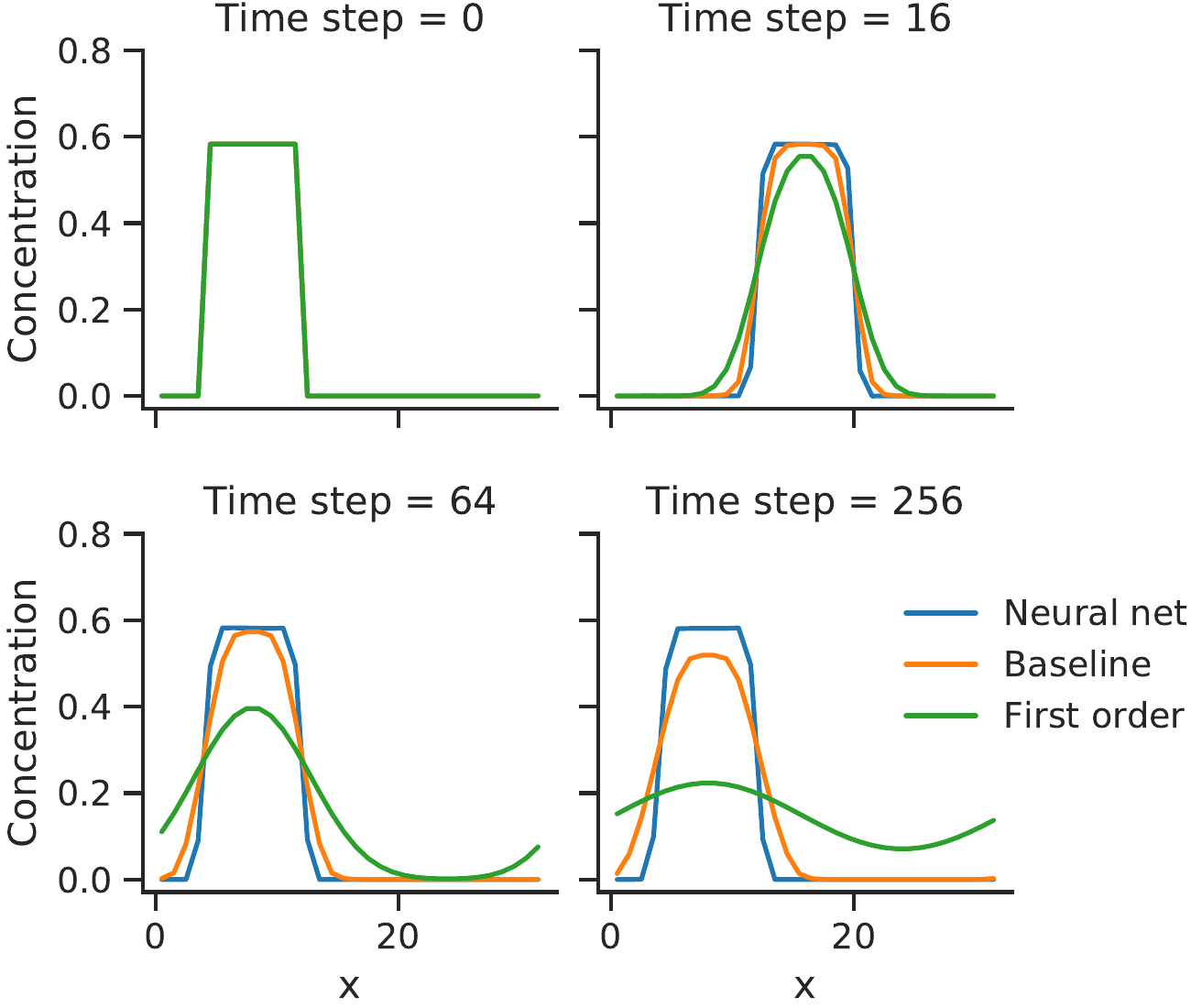}
  \caption{
  \textbf{One test sample for 1-D advection under constant velocity.} The concentration field is advected by half grid box every time step, and returns to the original position after every 64 time steps because the domain is periodic. Our neural network model is able to maintain the initial shape indefinitely, while traditional solvers accumulates numerical diffusion over time.
  }
  \label{fig:1d-sample}
\end{figure}

Fig.~\ref{fig:1d-sample} shows one test sample during the forward integration. The first-order upwind scheme exhibits large numerical diffusion, due to its second-order spatial discretizaion error \cite{odman1997}. The second-order VanLeer scheme (our baseline) is more accurate but stills accumulates diffusion over time. In contrast, our neural network model closely tracks the reference ``true solution" obtained by the $8\times$ resolution baseline. When a slight numerical diffusion occurs at one step, the next step applies a slight anti-diffusion to correct it. Intuitively, the solver learns that the optimal solution in one-dimensional advection is to maintain the initial shape.

\begin{figure}
  \centering
  \includegraphics[width=0.8\linewidth]{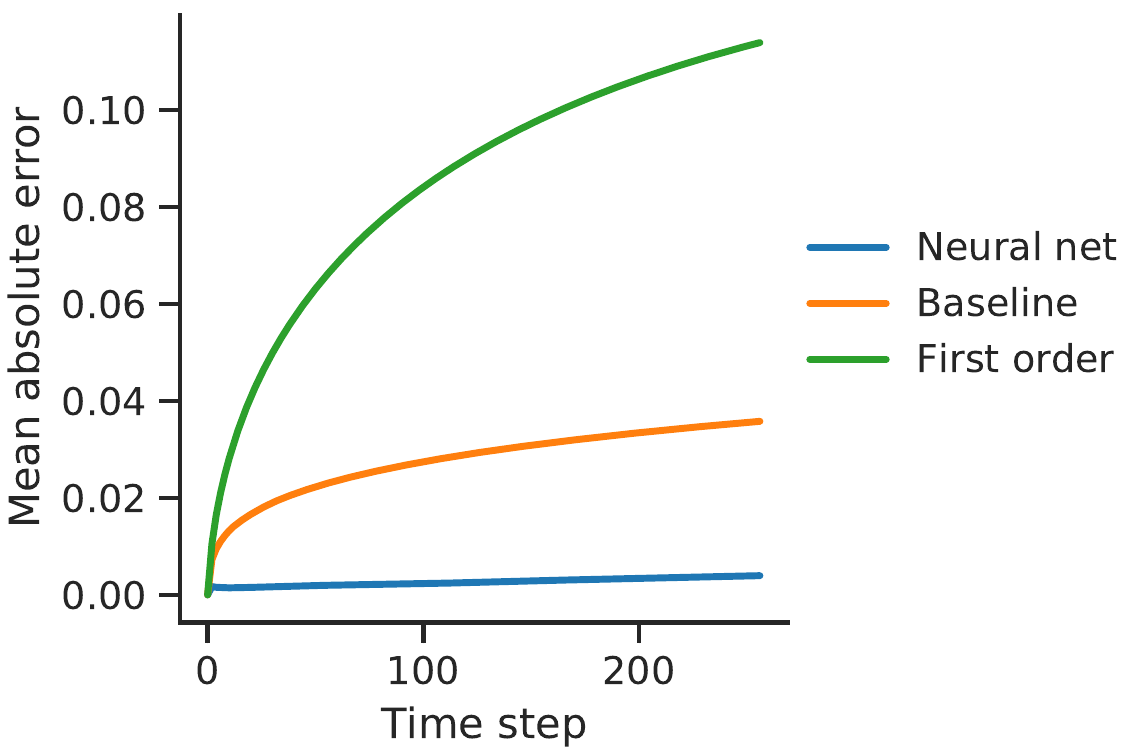}
  \caption{
  \textbf{Error for 1-D advection on test data.} Here we only plot even time steps (0, 2, 4, ...) for a smooth curve, because the error oscillates between odd and even time steps (a result of CFL=0.5).
  }
  \label{fig:1d-mae}
\end{figure}

Fig.~\ref{fig:1d-mae} shows the mean absolute error over time, averaged over all test samples. The error indicates the deviation from the reference solution obtained by the baseline solver at 256 grid points. The neural network model achieves a factor of 8 less error than the baseline second-order VanLeer scheme.

\begin{figure}
	\centering
	\includegraphics[width=1.0\linewidth]{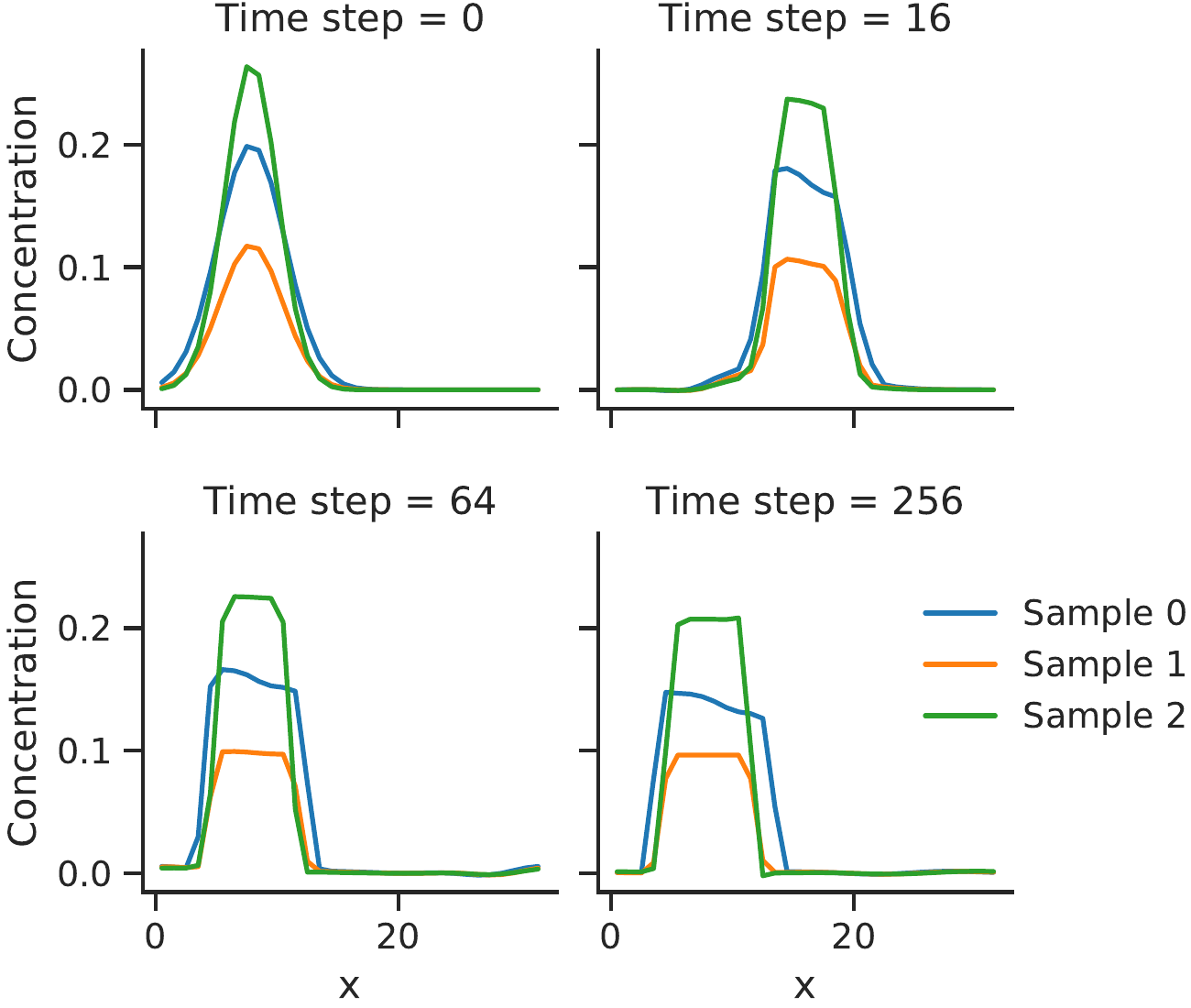}
	\caption{
	    \textbf{Neural network prediction on out-of-sample data.} The neural network model is only trained on square waves, but applied to Gaussian initial conditions. The model gradually turns Gaussian waves into squares, and then maintains the squares indefinitely.
	 }
	\label{fig:out-of-sample}
\end{figure}

We further investigate this intriguing behavior of our neural network model using out-of-sample test data. As shown in Fig.~\ref{fig:out-of-sample}, when the model (trained on square waves) is applied to Gaussian initial conditions, it gradually turns Gaussian waves into squares, which are the only shape in the training data. Then, the model can maintain the squares indefinitely. Such phenomenon of ``turning other shapes to squares" also exists in manually-designed schemes that are overly-optimized for square waves \cite{book2012}. The over-fitting problem here can be easily fixed by adding Gaussian shapes into training data; after that the neural network model can track both Gaussian and square shapes perfectly. Given that the input features for convolutional neural networks are localized in space, covering representative input patterns only requires limited amounts of training data.

\subsection{2-D deformational flow}

We next demonstrate that our neural network model can also achieve near-perfect result for a 2-D deformational flow test, originally proposed by \cite{leveque1996} and later extended to spherical coordinates as a standard test for atmospheric advection schemes \cite{nair2010,lauritzen2014}. The spatial domain is a square $[0, 1] \times [0,1]$, and the velocity field is a periodic swirling flow:
\begin{align}
  u(x,y,t) &= \sin^2(\pi x)\sin(2\pi y)\cos(\pi t/T) \\
  v(x,y,t) &= \sin^2(\pi y)\sin(2\pi x)\cos(\pi t/T) \nonumber
\end{align}
where the period $T=5$ in our setup. The direction of this flow reverses at $t=(n-\frac{1}{2})T$ for any positive integer $n$. The exact solution at $t=nT$ is identical to the initial condition.

The initial concentration field is a blob centered at $[1/4, 1/4]$:
\begin{align}
C(x,y) &= \frac{1}{2}[1 + \cos(\pi r)] \\
r(x,y) &= \min(1, 4\sqrt{(x-1/4)^2+(y-1/4)^2})
\nonumber
\end{align}

\begin{figure}
  \centering
  \includegraphics[width=\linewidth]{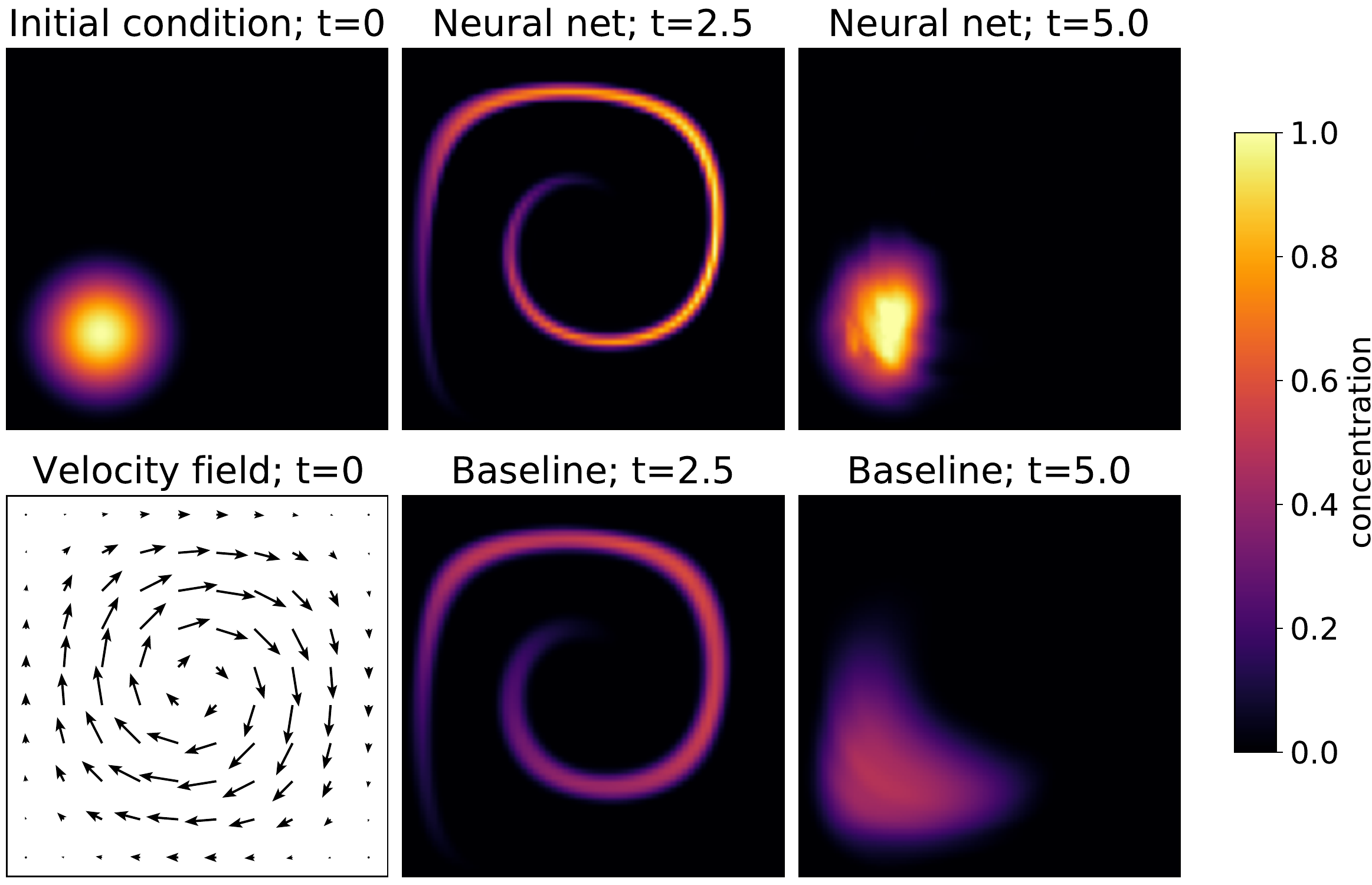}
  \caption{
  \textbf{Result on 2-D deformational flow.} The flow reverses at $t=2.5$ and returns to the initial condition at $t=5$. The neural network model is able to maintain a sharp gradient, while the baseline model incurs large numerical diffusion. The spatial domain is $[0, 1] \times [0, 1]$ (not plotted on axis).
  }
  \label{fig:2d-deform-mid}
\end{figure}

The model is not directly trained on this deformational flow, but instead on an ensemble of periodic, divergence-free, random velocity fields, implemented as superpositions of Sinusoidal waves as described by \cite{saad2016}. The trained model is able to generalize across different flows as long as the training data contain representative local patterns.

Fig.~\ref{fig:2d-deform-mid} shows the advection under deformational flow for the baseline and the neural network models, both on $64 \times 64$ grid points. The time step is chosen so that the maximum CFL number is 0.5. The neural network model is able to maintain a sharp concentration gradient, while the baseline VanLeer scheme incurs large numerical diffusion when the initial blob is stretched to a thin filament \cite{rastigejev2010}.

To quantify the numerical diffusion, we use the entropy $S$ as a metric \cite{zhuang2018}:
\begin{equation}
S = - \beta \sum_i C_i \cdot \log(C_i)
\label{eq:entropy}
\end{equation}
where the concentration $C$ is scaled such that the initial conditions falls in the range $[0, 1]$, and $\beta$ is a normalization factor so that the initial entropy is 1.
Entropy is conserved under pure advection and increases under diffusion.
To avoid an undefined answer if any $C_i < 0$, we use first set negative values in the concentration to zero, and evaluate $C=0$ via the limit $x \log x = 0$ as $x \to 0$.

\begin{figure}
  \centering
  \includegraphics[width=1.0\linewidth]{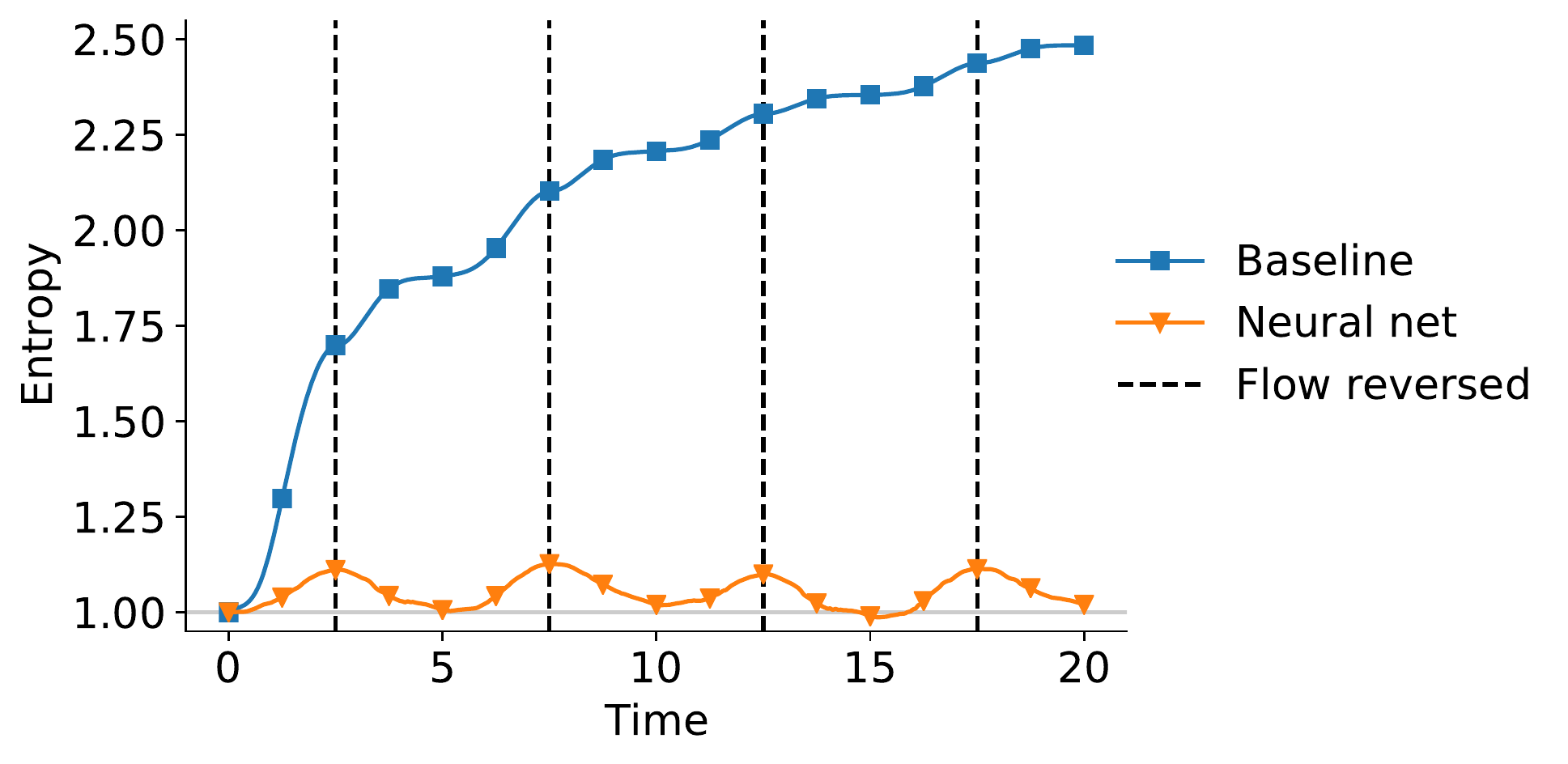}
  \caption{
  \textbf{Entropy for advection under 2-D deformational flow.} Entropy is conserved under pure advection and increases under diffusion. Traditional monotonic solvers are only allowed to increase entropy, while our neural network model is allowed to decrease entropy and and thus minimizes diffusion error over a long time.
  }
  \label{fig:2d-deform-entropy}
\end{figure}

Fig.~\ref{fig:2d-deform-entropy} shows the entropy over time. Any monotonic advection solver can only increase entropy; any entropy decrease indicates nonphysical anti-diffusion, which often occurs due to numerical instability. Strikingly, the neural network model {\sl can} decrease entropy, while still remaining numerically stable. Although such behavior seems to be nonphysical, it is indeed the best possible solution on such a coarse grid. On a grid that perfectly resolves the concentration field, the entropy remains constant under the deformational flow. Yet on a coarse grid view, the computed entropy increases when the initial blob turns into filament due to conservative averaging, and then decreases when the filament reverts back into a blob. Our neural network model can disobey the commonly-used constraint of non-decreasing entropy, and thus more closely matches the exact solution, when compared to traditional monotonic solvers.

\subsection{2-D turbulent flow} \label{section:turbulence}

As the final test, we use the velocity fields from freely-evolving, decaying 2-D turbulence simulations in pyqg (\url{https://github.com/pyqg/pyqg}). The spatial domain is $[0, 2\pi] \times [0, 2\pi]$ with periodic boundary condition. We use a $256 \times 256$ grid for generating the reference solution using the baseline solver, and a $32 \times 32$ coarse grid for model evaluation. As in previous cases, here the advection time step is chosen so that the maximum CFL number is 0.5. 

The training and test velocities are generated from different random seeds. We start with the McWilliams-84 random initial condition \cite{mcwilliams1984} and let the turbulence decay with time. We discard the initial 4 seconds of the simulation so that the velocity field can be resolved on the coarse grid. For the initial concentration field, we use an ensemble of 10 blobs with width 0.5 at random locations. Note that the spatial scale of the concentration field under turbulent advection can become much smaller than the scale of the velocity field \cite{shraiman2000,methven1999}, making it challenging for traditional advection solvers to resolve the concentration gradient. We use 20 random initial conditions for training data and 20 for test data. The actual sample size for the training dataset is much larger, as each initial condition is integrated into a time series of 1024 steps on the fine grid or 128 steps on the coarse grid, which is further broken into many 10-step time series for calculating the multi-step loss function.

\begin{figure*}
  \centering
  \includegraphics[width=\linewidth]{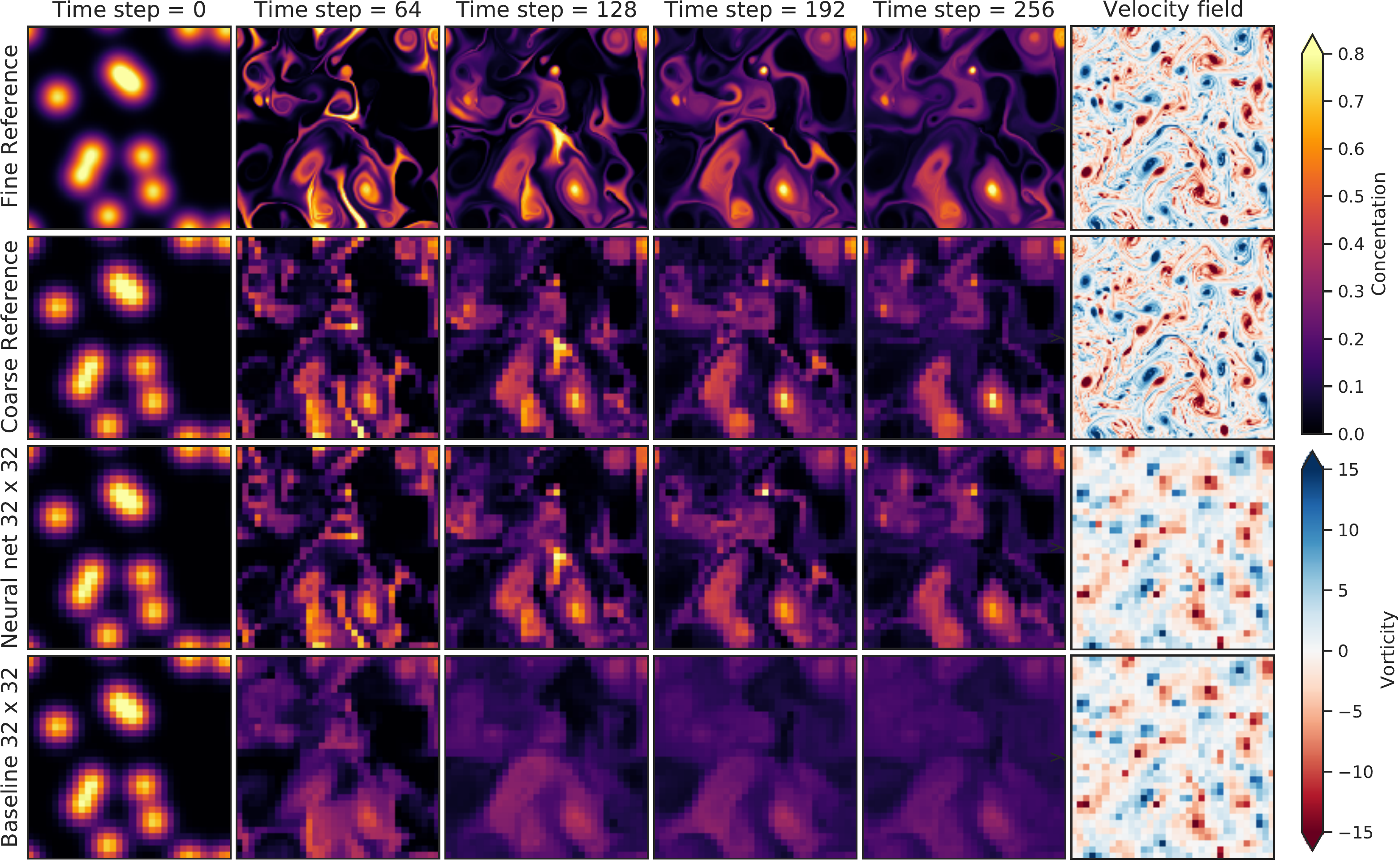}
  \caption{
  \textbf{One test sample under 2-D turbulent flow.} The initial blobs (first column) are stretched into thin filaments under the turbulent flow (last column), illustrated by the vorticity $\omega = \partial_x u_y - \partial_y u_v$. The baseline solver (second-order VanLeer scheme) can resolve such filaments on the fine-resolution grid, but incurs large numerical diffusion on the coarse grid. The neural network model can preserve the sharp features on the coarse grid. The spatial domain is $[0, 2\pi] \times [0, 2\pi]$ (not plotted on axis). 
  }
  \label{fig:turbulent-result}
\end{figure*}

Fig.~\ref{fig:turbulent-result} shows one test sample under the 2-D turbulent flow, for both the initial condition (the left column of Fig.~\ref{fig:turbulent-result}) and the integration results after 256 time steps (the middle and right columns of Fig.~\ref{fig:turbulent-result}). Note that this is twice the maximum number of time-steps used for model training. The initial blobs are stretched into thin filaments under the turbulent flow. The baseline solver (second-order VanLeer scheme) can resolve such filaments on the fine-resolution grid, but incurs large numerical diffusion on the coarse grid and loses the sharp concentration gradient. However, when the fine-grid solution is directly resampled to the coarse grid, most sharp features can actually be preserved. Thus, the inability to resolve sharp gradients is not due to the coarse grid itself, but instead due to the numerical error in the baseline advection solver. Our neural network model, trained to track the optimal reference solution on the coarse grid, is able to preserve sharp features during the forward integration. The model performs well on all test samples, with more shown in Appendix.

\begin{figure}
  \centering
  \includegraphics[width=1.0\linewidth]{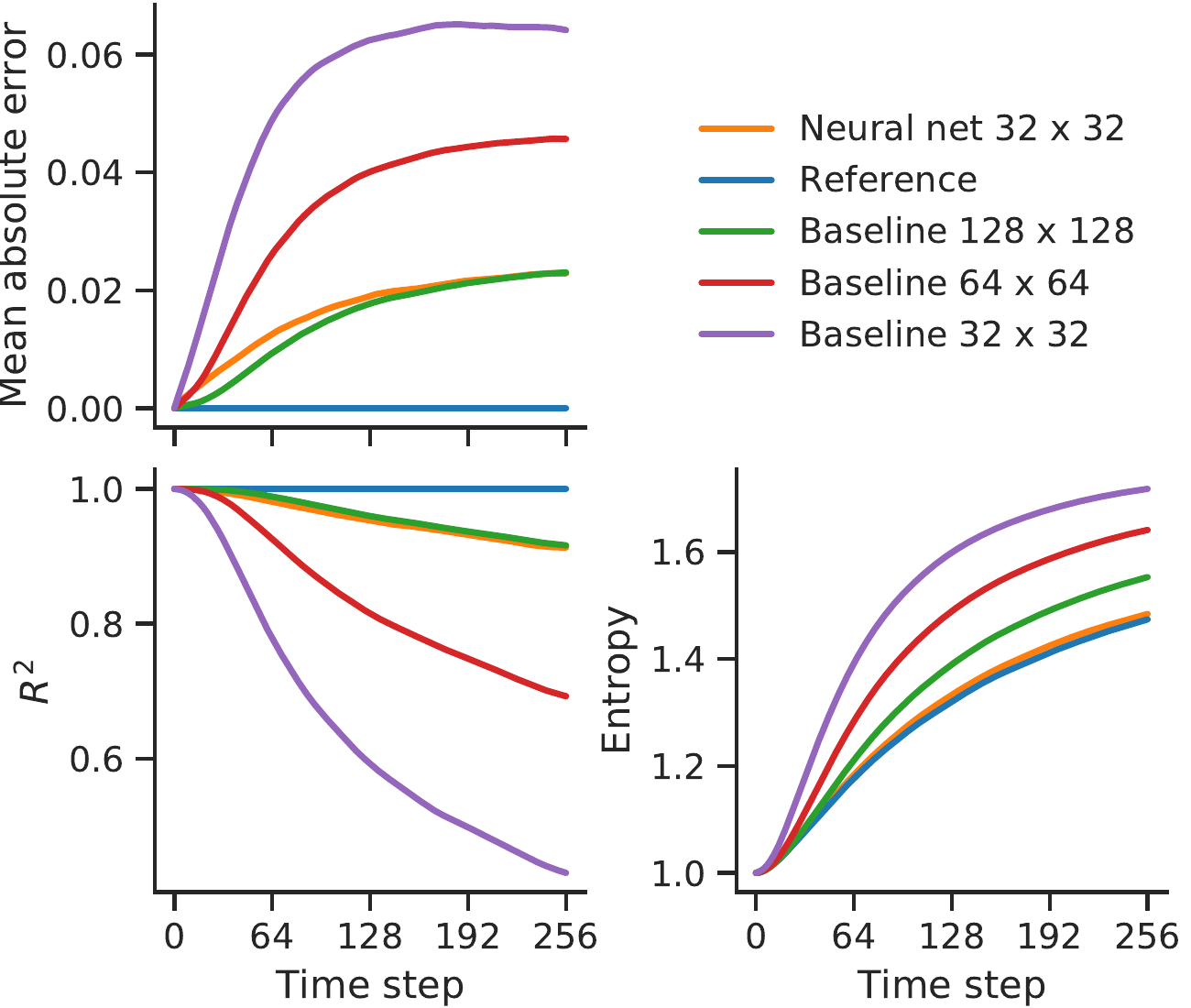}
  \caption{
  \textbf{Error for 2-D turbulent advection on test data.} The neural network model achieves the same accuracy as the second-order VanLeer baseline scheme at $4\times$ resolution, and entropy similar to the baseline at $8\times$ resolution.
  }
  \label{fig:turbulent-all-metrics}
\end{figure}

Fig.~\ref{fig:turbulent-all-metrics} shows a variety of error metrics for advection under turbulent flow, averaged over all test samples.
The error is computed as the deviation from the reference solution obtained by the baseline solver at $256 \times 256$ grid. We also compare the baseline solver at intermediate grid resolutions ($64 \times 64$ and $128 \times 128$). All solutions are resampled to the $32 \times 32$ coarse grid for error calculation.
We use two measurements of accuracy, mean absolute error (our training loss) and the coefficient of determination $R^2$, which means the goodness of fit for linear regression models for to reference solution.
Based these metrics, our neural network model achieves roughly the same accuracy as the baseline method at $4\times$ resolution ($128 \times 128$).
We also evaluate the entropy for all solutions based on Eq.~\eqref{eq:entropy}, which suggests that from a statistical perspective our neural net model almost perfectly matches the reference simulation on which it was trained.

\begin{figure}
  \centering
  \includegraphics[width=1.0\linewidth]{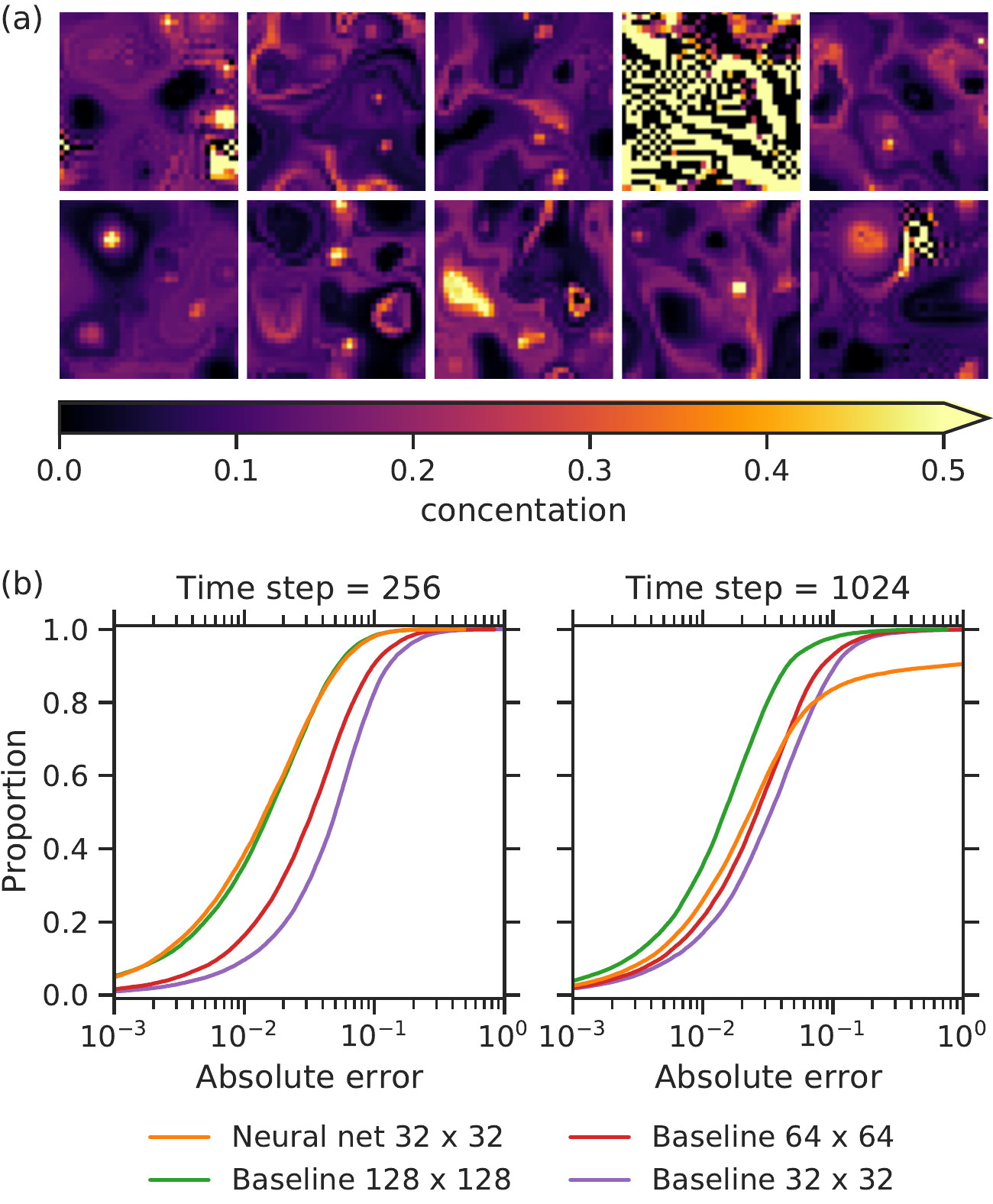}
  \caption{
  \textbf{Limitations of long time stability under 2-D turbulent flow.}
    (a) Ten randomly chosen examples of concentrations fields from the neural net model after 1024 time steps. One field (first row, fourth column) is entirely covered by ``checkerboard'' artifacts, and two others (top left and bottom right) show checkerboard artifacts in limited regions.
    (b) Empirical distribution function for absolute error across all models after 256 and 1024 time steps. The neural net performance suffers significantly, with about 10\% of solutions having an absolute error greater than 1.
  }
  \label{fig:turbulent-long-time}
\end{figure}

Figure~\ref{fig:turbulent-long-time} illustrates the limitations of stability and generalization for our neural net model by integrating for far longer than the 128 time steps used for training data.
After 1000 time integration steps, our neural net model shows obvious numerical artifacts (checkerboard patterns) and very poor accuracy for about 10\% of random seeds.
Unlike the baseline models, our neural net model does not guarantee properties such as monotonicity, and when presented with examples outside of its training data it occasionally extrapolates poorly.
Figuring out how to guarantee stability for neural net models, either by training on more comprehensive datasets or by imposing architecture constraints, is an important topic for future work.

\begin{figure}
  \centering
  \includegraphics[width=1.0\linewidth]{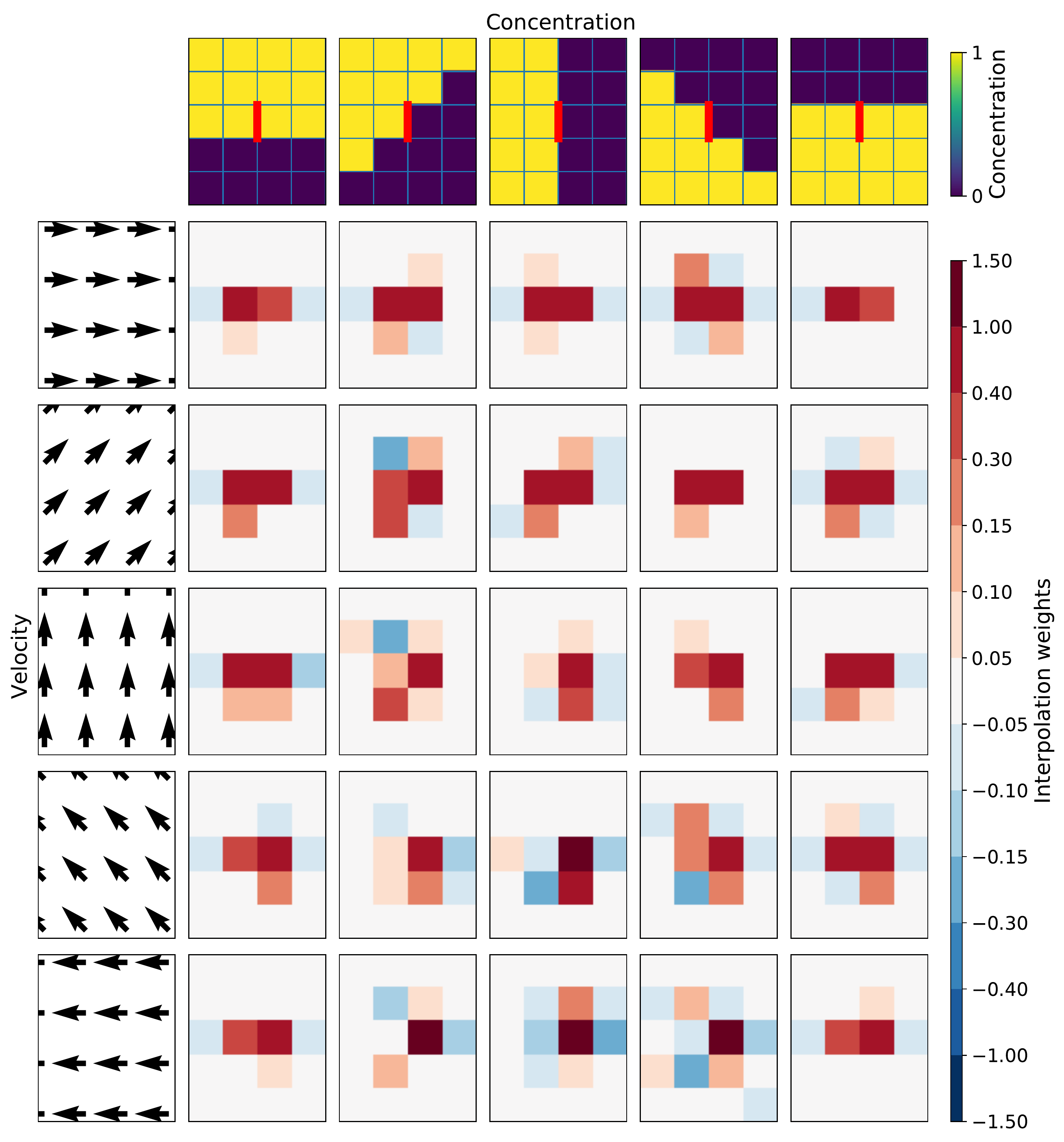}
  \caption{
  \textbf{Interpolation coefficients for 2D advection.} Illustration of how prediction of the interpolation coefficients changes for different combinations of concentration (top row) and velocity field (left column) inputs. Concentration values represented by color; velocity field has unit magnitude and changes direction as shown in the plot. The target location for the interpolation is marked by a red bar on the concentration plots. The model predominantly interpolates along the velocity field and concentration edges, rediscovering the upwinding-like methods at the corner cases of the facet. While the model learned some general symmetries, we expect even better results for models that incorporate symmetries by construction.
  }
  \label{fig:2d-filters}
\end{figure}

Finally, to get a glimpse into the inner workings of the trained model, Fig.~\ref{fig:2d-filters} examines predicted interpolation coefficients for $x$ component of the velocity field. We see that similar to hand-crafted methods, the learned interpolation depends on both velocity and concentration. While some of the symmetries have been clearly learned from the data, we believe that incorporating such priors could improve the results further.

\section{Computational performance and accuracy with different hyperparameters} \label{section:speed}

\begin{figure}
  \centering
  \includegraphics[width=0.9\linewidth]{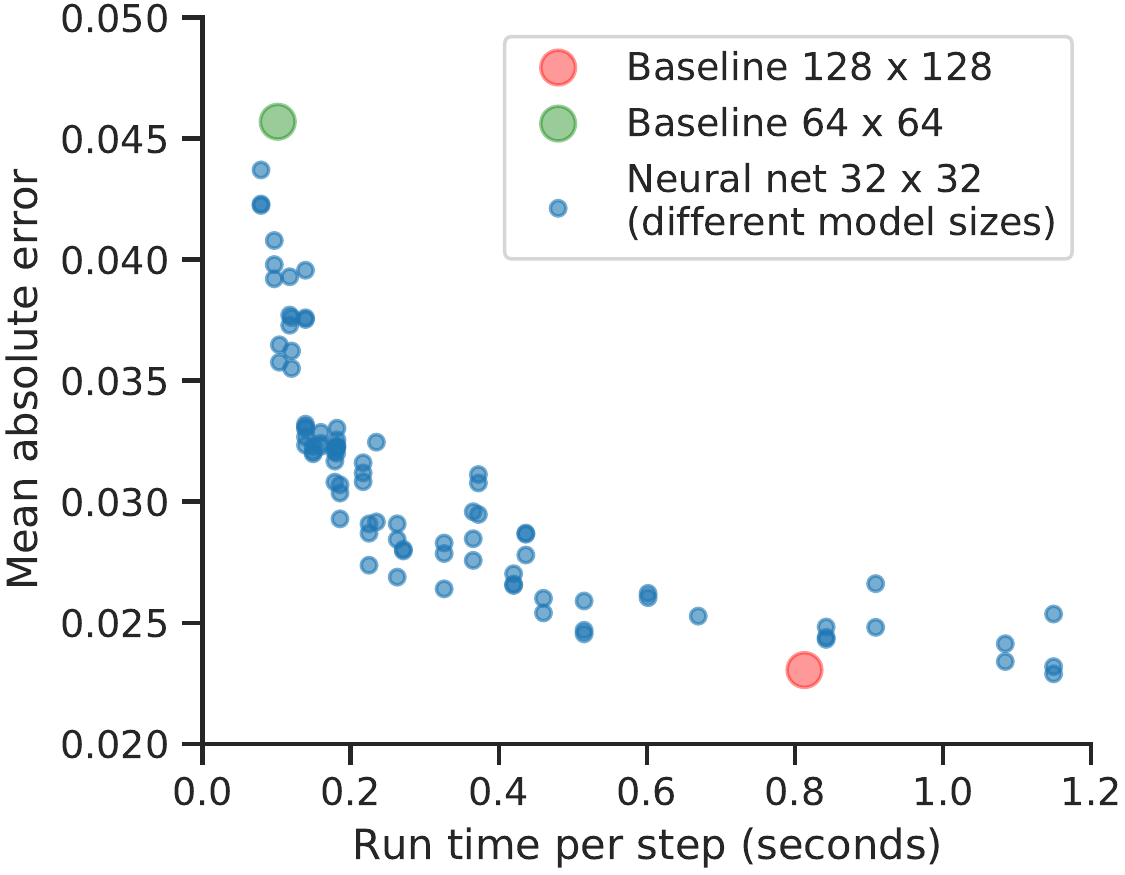}
  \caption{
  \textbf{Accuracy-speed tradeoff for neural network model.}
  Each data point is a neural network model with different hyperparamters (detailed in Fig.~\ref{fig:tune-hyperparam}). The performance of the baseline solver at intermediate grid resolutions ($64 \times 64$ and $128 \times 128$) is overlaid for comparison. The model accuracy is evaluated on the 2-D turbulence case after 256 time steps (Section \ref{section:turbulence}), and the run time is measured on a single Nvidia V100 GPU. The x-axis shows the wall-clock time per advection time step on the coarse grid, which requires 2 or 4 time steps for the $64 \times 64$ or $128 \times 128$ baseline due to the CFL condition.
  }
  \label{fig:accuracy-speed}
\end{figure}

\begin{figure}
	\centering
	\includegraphics[width=\linewidth]{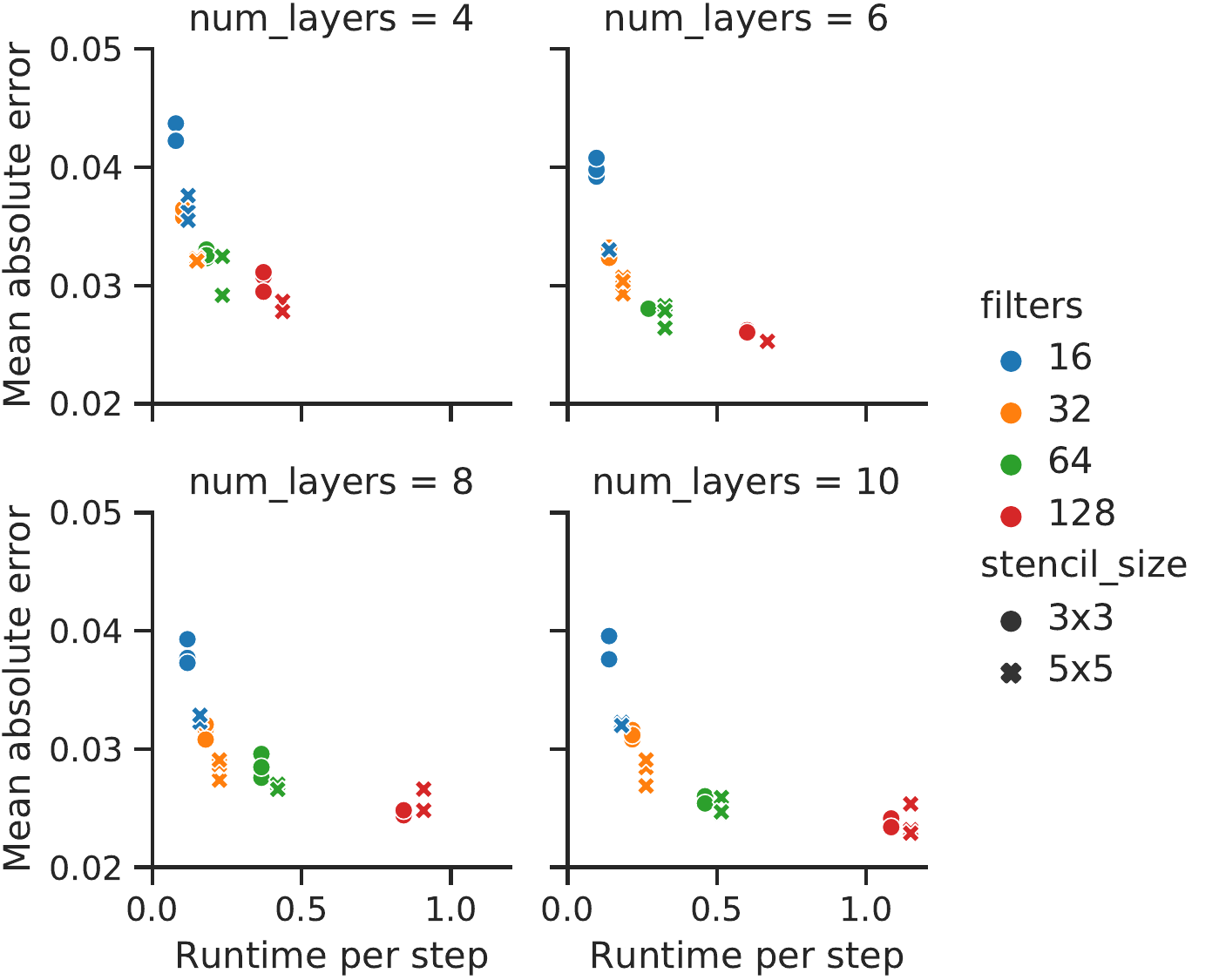}
	\caption{
	    \textbf{Hyper-parameter effect on neural network model performance.} Same as Fig.~\ref{fig:accuracy-speed}, but here the data points are grouped by different numbers of convolutional layers and the number of filters, with shape denoted the size of the stencil. Duplicate points corresponds to identical models trained with different random seeds.
	 }
	\label{fig:tune-hyperparam}
\end{figure}

There is a tradeoff between accuracy and speed for our neural network model, as using a larger convolutional neural network increases both the accuracy and the run time. We performed a grid search on model hyperparameters, for the number of layers ranging from [4, 6, 8, 10], the number of convolutional filers ranging from [16, 32, 64, 128], and the finite-difference stencil size ranging from [3, 5], with each case replicated 3 times with different random seeds. The model accuracy is evaluated on the 2-D turbulence case in Section \ref{section:turbulence}, and the run time is measured on a single Nvidia V100 GPU.

Fig.~\ref{fig:accuracy-speed} shows the model accuracy and speed using different hyperparameters. The performance of the baseline solver at intermediate grid resolutions ($64 \times 64$ and $128 \times 128$) is overlaid on for comparison. A large neural network ($8 \sim 10$ layers and 128 filters) achieves comparable accuracy and speed as the baseline solver at $4 \times$ resolution, while a small neural network (4 layers and 32 filters) performs similarly to the baseline solver at $2 \times$ resolution. Fig.~\ref{fig:tune-hyperparam} shows that using 64 filters and 10 layers achieves a good balance between accuracy and speed, in which case the model achieves similar accuracy as the $4 \times$ resolution baseline while being $80\%$ faster.

The model speed largely depends on the code implementation and software configuration. Our current implementation of the neural network model has a lot of room for performance optimization. For example, our code still requires unnecessary, large memory allocation, and does not use the reduced-precision Tensor Cores in the V100 GPU. With more performance tuning as well as techniques like model compression and quantization \cite{cheng2017}, the neural network model may significantly outperform the baseline in terms of computational performance.

Incorporating neural networks into numerical methods also allows better utilization of current and emerging hardware. It is reported that ``current (Earth system) models running on next generation architectures use less than 2\% of peak performance" \cite{carman2017}. This is because classic numerical methods (e.g. finite difference, finite volume) are limited by memory bandwidth rather than processor speed \cite{schulthess2018} \footnote{Indeed, not all traditional solvers have ultra-low machine utilization. For example, the discontinuous Galerkin (DG) method can have a much higher machine utilization of 20\% \cite{breuer2019}. Although computational fluid dynamics (CFD) models can exploit high-order methods, climate models tend to use low-order schemes \cite{schulthess2018}.}. In contrast, neural networks mostly consist of dense matrix multiplications with a high compute-to-memory ratio, and therefore can achieve near-peak performance on both CPUs and hardware accelerators (see the Roofline charts \cite{Williams2009} in \cite{jouppi2017}). We measure the machine utilization using Perf on CPU and NVProf on GPU, and find that the neural network model achieves $80\%$ of peak FLOPs (floating point operations per second), while the baseline solver only uses $1 \sim 2\%$ of peak FLOPs. Clever use of neural network emulations inside existing models may provide a way to squeeze out ``free compute cycles" that are currently not utilized.

\section{Conclusion}

We developed a data-driven discretization for solving passive scalar advection in one- or two- dimensions. Instead of using pre-defined coefficients to compute the spatial derivatives in the partial differential equation (PDE), we used a convolutional neural network to learn the optimal finite difference coefficients, so that the solution best matches the ``true" result obtained by high-resolution reference simulations. 

Our neural-network-based model is able to obtain near-perfect results for idealized 1-D and 2-D test problems, while a traditional high-order solver incurs significant diffusion error. Under a 2-D turbulent flow, the neural network model running on a coarse grid can achieve the same accuracy as a traditional high-order solver running on a $4\times$ higher resolution grid.

The neural network model exhibits several interesting behaviors that may help explain its unusual accuracy. Our learned models have been specifically optimized for modeling specific class of flows used as training data, which limits their range of validity. For example, in 1D the model converts unseen shapes into known shapes, and on 2D turbulent flows the model occasionally fails entirely when asked to make predictions for much longer times than were used in training. An important challenge for future work is identify techniques that can ensure learned discretizations are robust even to such out-of-distribution inputs. Alternatively, it may be able to identify training datasets that cover the full range of phenomena of interest, e.g., in the context of weather or pollution forecasts where the same equations are solved day after day.

At the same accuracy, the speed of our neural network model is comparable to the baseline high-order solver (that runs at $4 \times$ higher resolution). There is a lot of room for further optimizing the neural network performance in our code implementation. Notably, the neural network model can achieve a much higher machine utilization than traditional finite-difference methods, and will better utilize emerging hardware accelerators.

An open question is how to apply our method in existing computational fluid dynamics (CFD) or climate/weather models, which tend to be implemented in large codebases of C++ or Fortran. Although past work has successfully replaced one component in a model with neural networks \cite{rasp2018}, our approach works best in an end-to-end differentiable program. Recent efforts in implementing models in Julia \cite{clima} and JAX \cite{schoenholz2019} should ease the integration of scientific computing and machine learning. 

Code and tutorials for this work are available at \url{https://github.com/google-research/data-driven-pdes}.

\section*{Acknowledgements}

We thank Anton Gerashenko, Jamie Smith, Peynman Milanfar, Pascal
Getreur, Ignacio Garcia Dorado and Rif Saurous for important conversations.
Y.B.S acknowledges support from the James S. McDonnel
post-doctoral fellowship for the study of complex systems.
M.P.B acknowledges support from NSF DMS-1715477 as well
as the Simons Foundation.

\renewcommand{\thefigure}{S\arabic{figure}}
\renewcommand{\theequation}{S-\arabic{equation}}
\setcounter{figure}{0}
\setcounter{equation}{0}

\section*{Appendix: Sample results for 2-D turbulent advection}

\begin{figure*}
	\centering
	\includegraphics[width=1.0\linewidth]{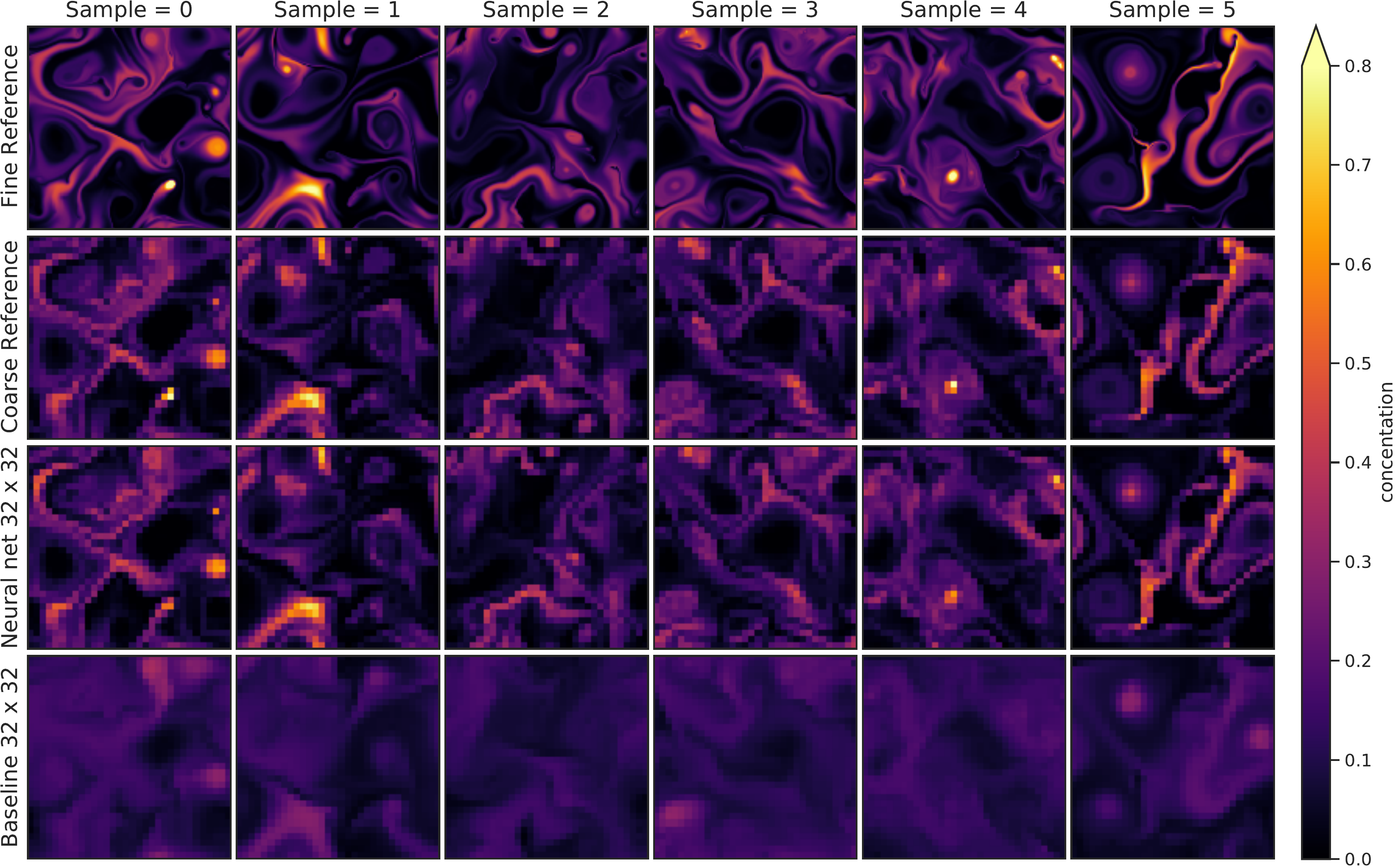}
	\caption{
	    \textbf{Sample results for advection under 2-D turbulent flow.}
	    Concentration fields after 256 time steps are illustrated for six different randomly initialized concentration and velocity fields. The models are the same as those illustrated in Fig.~\ref{fig:turbulent-result}. 
	 }
	\label{fig:turbulent-samples}
\end{figure*}

Figure \ref{fig:turbulent-samples} shows more test samples for the 2-D turbulent advection problem in Section \ref{section:turbulence}. In all test samples, the neural network model is able to maintain a sharp gradient that closely matches the reference true resolution, while the baseline model incurs significant numerical diffusion error.

\clearpage  

\bibliography{mybibfile}

\begin{thebibliography}{10}
\expandafter\ifx\csname url\endcsname\relax
  \def\url#1{\texttt{#1}}\fi
\expandafter\ifx\csname urlprefix\endcsname\relax\def\urlprefix{URL }\fi
\expandafter\ifx\csname href\endcsname\relax
  \def\href#1#2{#2} \def\path#1{#1}\fi

\bibitem{jimenez2019}
J.~Jim{\'e}nez, Computers and turbulence, European Journal of
  Mechanics-B/Fluids (2019).

\bibitem{bauer2015}
P.~Bauer, A.~Thorpe, G.~Brunet, The quiet revolution of numerical weather
  prediction, Nature 525~(7567) (2015) 47--55 (2015).

\bibitem{schneider2017a}
T.~Schneider, J.~Teixeira, C.~S. Bretherton, F.~Brient, K.~G. Pressel,
  C.~Sch{\"a}r, A.~P. Siebesma, Climate goals and computing the future of
  clouds, Nature Climate Change 7~(1) (2017) 3--5 (2017).

\bibitem{neumann2019}
P.~Neumann, P.~D{\"u}ben, P.~Adamidis, P.~Bauer, M.~Br{\"u}ck, L.~Kornblueh,
  D.~Klocke, B.~Stevens, N.~Wedi, J.~Biercamp, Assessing the scales in
  numerical weather and climate predictions: will exascale be the rescue?,
  Philosophical Transactions of the Royal Society A 377~(2142) (2019) 20180148
  (2019).

\bibitem{theis2017}
T.~N. Theis, H.-S.~P. Wong, The end of moore's law: A new beginning for
  information technology, Computing in Science \& Engineering 19~(2) (2017)
  41--50 (2017).

\bibitem{shalf2020}
J.~Shalf, The future of computing beyond moore’s law, Philosophical
  Transactions of the Royal Society A 378~(2166) (2020) 20190061 (2020).

\bibitem{schneider2017b}
T.~Schneider, S.~Lan, A.~Stuart, J.~Teixeira, Earth system modeling 2.0: A
  blueprint for models that learn from observations and targeted
  high-resolution simulations, Geophysical Research Letters 44~(24) (2017)
  12--396 (2017).

\bibitem{reichstein2019}
M.~Reichstein, G.~Camps-Valls, B.~Stevens, M.~Jung, J.~Denzler, N.~Carvalhais,
  et~al., Deep learning and process understanding for data-driven earth system
  science, Nature 566~(7743) (2019) 195--204 (2019).

\bibitem{kutz2017}
J.~N. Kutz, Deep learning in fluid dynamics, Journal of Fluid Mechanics 814
  (2017) 1--4 (2017).

\bibitem{duraisamy2019}
K.~Duraisamy, G.~Iaccarino, H.~Xiao, Turbulence modeling in the age of data,
  Annual Review of Fluid Mechanics 51 (2019) 357--377 (2019).

\bibitem{hatfield2019}
S.~Hatfield, M.~Chantry, P.~D{\"u}ben, T.~Palmer, Accelerating high-resolution
  weather models with deep-learning hardware, in: Proceedings of the Platform
  for Advanced Scientific Computing Conference, 2019, pp. 1--11 (2019).

\bibitem{hennessy2019}
J.~L. Hennessy, D.~A. Patterson, A new golden age for computer architecture,
  Communications of the ACM 62~(2) (2019) 48--60 (2019).

\bibitem{dean2019}
J.~Dean, The deep learning revolution and its implications for computer
  architecture and chip design, arXiv preprint arXiv:1911.05289 (2019).

\bibitem{bar2019}
Y.~Bar-Sinai, S.~Hoyer, J.~Hickey, M.~P. Brenner, Learning data-driven
  discretizations for partial differential equations, Proceedings of the
  National Academy of Sciences 116~(31) (2019) 15344--15349 (2019).

\bibitem{shraiman2000}
B.~I. Shraiman, E.~D. Siggia, Scalar turbulence, Nature 405~(6787) (2000)
  639--646 (2000).

\bibitem{rastigejev2010}
Y.~Rastigejev, R.~Park, M.~P. Brenner, D.~J. Jacob, Resolving intercontinental
  pollution plumes in global models of atmospheric transport, Journal of
  Geophysical Research: Atmospheres 115~(D2) (2010).

\bibitem{leveque1996}
R.~J. LeVeque, High-resolution conservative algorithms for advection in
  incompressible flow, SIAM Journal on Numerical Analysis 33~(2) (1996)
  627--665 (1996).

\bibitem{sweby1984}
P.~K. Sweby, High resolution schemes using flux limiters for hyperbolic
  conservation laws, SIAM journal on numerical analysis 21~(5) (1984) 995--1011
  (1984).

\bibitem{lin1996}
S.-J. Lin, R.~B. Rood, Multidimensional flux-form semi-lagrangian transport
  schemes, Monthly Weather Review 124~(9) (1996) 2046--2070 (1996).

\bibitem{ullrich2010}
P.~A. Ullrich, C.~Jablonowski, B.~Van~Leer, High-order finite-volume methods
  for the shallow-water equations on the sphere, Journal of Computational
  Physics 229~(17) (2010) 6104--6134 (2010).

\bibitem{colella2008}
P.~Colella, M.~D. Sekora, A limiter for ppm that preserves accuracy at smooth
  extrema, Journal of Computational Physics 227~(15) (2008) 7069--7076 (2008).

\bibitem{semakin2016}
A.~Semakin, Y.~Rastigejev, Numerical simulation of global-scale atmospheric
  chemical transport with high-order wavelet-based adaptive mesh refinement
  algorithm, Monthly Weather Review 144~(4) (2016) 1469--1486 (2016).

\bibitem{stohl2005}
A.~Stohl, C.~Forster, A.~Frank, P.~Seibert, G.~Wotawa, The lagrangian particle
  dispersion model flexpart version 6.2 (2005).

\bibitem{eastham2017}
S.~D. Eastham, D.~J. Jacob, Limits on the ability of global eulerian models to
  resolve intercontinental transport of chemical plumes., Atmospheric Chemistry
  \& Physics 17~(4) (2017).

\bibitem{lauritzen2010}
P.~H. Lauritzen, R.~D. Nair, P.~A. Ullrich, A conservative semi-lagrangian
  multi-tracer transport scheme (cslam) on the cubed-sphere grid, Journal of
  Computational Physics 229~(5) (2010) 1401--1424 (2010).

\bibitem{kulkarni2019}
C.~S. Kulkarni, P.~F. Lermusiaux, Advection without compounding errors through
  flow map composition, Journal of Computational Physics 398 (2019) 108859
  (2019).

\bibitem{Note1}
Using this MAE loss indicates that we require point-wise agreement between the
  machine learning prediction and the reference true solution, on every time
  step. This criteria is relevant for atmospheric transport modeling where the
  goal is to simulate the instantaneous, deterministic concentration field
  \cite {rastigejev2010}. For turbulence modeling, matching the high-order
  statistics might be more desirable than requiring point-wise agreement of
  instantaneous fields. This would require changes in the loss function.

\bibitem{brenowitz2018}
N.~D. Brenowitz, C.~S. Bretherton, Prognostic validation of a neural network
  unified physics parameterization, Geophysical Research Letters 45~(12) (2018)
  6289--6298 (2018).

\bibitem{innes2019}
M.~Innes, A.~Edelman, K.~Fischer, C.~Rackauckus, E.~Saba, V.~B. Shah,
  W.~Tebbutt, Zygote: A differentiable programming system to bridge machine
  learning and scientific computing, arXiv preprint arXiv:1907.07587 (2019).

\bibitem{baydin2017}
A.~G. Baydin, B.~A. Pearlmutter, A.~A. Radul, J.~M. Siskind, Automatic
  differentiation in machine learning: a survey, The Journal of Machine
  Learning Research 18~(1) (2017) 5595--5637 (2017).

\bibitem{bischof1996}
C.~Bischof, P.~Khademi, A.~Mauer, A.~Carle, Adifor 2.0: Automatic
  differentiation of fortran 77 programs, IEEE Computational Science and
  Engineering 3~(3) (1996) 18--32 (1996).

\bibitem{abadi2016}
M.~Abadi, P.~Barham, J.~Chen, Z.~Chen, A.~Davis, J.~Dean, M.~Devin,
  S.~Ghemawat, G.~Irving, M.~Isard, et~al., Tensorflow: A system for
  large-scale machine learning, in: 12th $\{$USENIX$\}$ Symposium on Operating
  Systems Design and Implementation ($\{$OSDI$\}$ 16), 2016, pp. 265--283
  (2016).

\bibitem{paszke2017}
A.~Paszke, S.~Gross, S.~Chintala, G.~Chanan, E.~Yang, Z.~DeVito, Z.~Lin,
  A.~Desmaison, L.~Antiga, A.~Lerer, Automatic differentiation in pytorch
  (2017).

\bibitem{jax}
J.~Bradbury, R.~Frostig, P.~Hawkins, M.~J. Johnson, C.~Leary, D.~Maclaurin,
  S.~Wanderman-Milne, Jax: composable transformations of python+ numpy
  programs, URL http://github.com/google/jax.

\bibitem{innes2018}
M.~Innes, Flux: Elegant machine learning with julia, Journal of Open Source
  Software 3~(25) (2018) 602 (2018).

\bibitem{swift}
Swift differentiable programming manifesto, 2018, URL
  https://github.com/apple/swift.

\bibitem{schoenholz2019}
S.~S. Schoenholz, E.~D. Cubuk, Jax, md: End-to-end differentiable, hardware
  accelerated, molecular dynamics in pure python, arXiv preprint
  arXiv:1912.04232 (2019).

\bibitem{agrawal2019}
A.~Agrawal, A.~N. Modi, A.~Passos, A.~Lavoie, A.~Agarwal, A.~Shankar,
  I.~Ganichev, J.~Levenberg, M.~Hong, R.~Monga, et~al., Tensorflow eager: A
  multi-stage, python-embedded dsl for machine learning, arXiv preprint
  arXiv:1903.01855 (2019).

\bibitem{lin1994}
S.-J. Lin, W.~C. Chao, Y.~Sud, G.~Walker, A class of the van leer-type
  transport schemes and its application to the moisture transport in a general
  circulation model, Monthly Weather Review 122~(7) (1994) 1575--1593 (1994).

\bibitem{smaoui2001}
H.~Smaoui, B.~Radi, Comparative study of different advective schemes:
  Application to the mecca model, Environmental Fluid Mechanics 1~(4) (2001)
  361--381 (2001).

\bibitem{hassanzadeh2009}
H.~Hassanzadeh, J.~Abedi, M.~Pooladi-Darvish, A comparative study of
  flux-limiting methods for numerical simulation of gas--solid reactions with
  arrhenius type reaction kinetics, Computers \& Chemical Engineering 33~(1)
  (2009) 133--143 (2009).

\bibitem{raissi2018}
M.~Raissi, G.~E. Karniadakis, Hidden physics models: Machine learning of
  nonlinear partial differential equations, Journal of Computational Physics
  357 (2018) 125--141 (2018).

\bibitem{raissi2019}
M.~Raissi, P.~Perdikaris, G.~E. Karniadakis, Physics-informed neural networks:
  A deep learning framework for solving forward and inverse problems involving
  nonlinear partial differential equations, Journal of Computational Physics
  378 (2019) 686--707 (2019).

\bibitem{Bezenac2019-ia}
E.~d. B{\'e}zenac, E.~de~B{\'e}zenac, A.~Pajot, P.~Gallinari,
  \href{http://dx.doi.org/10.1088/1742-5468/ab3195}{Deep learning for physical
  processes: incorporating prior scientific knowledge} (2019).
\newline\urlprefix\url{http://dx.doi.org/10.1088/1742-5468/ab3195}

\bibitem{Um2020-zf}
K.~Um, {Raymond}, {Fei}, P.~Holl, R.~Brand, N.~Thuerey,
  \href{http://arxiv.org/abs/2007.00016}{{Solver-in-the-Loop}: Learning from
  differentiable physics to interact with iterative {PDE-Solvers}} (Jun. 2020).
\newblock \href {http://arxiv.org/abs/2007.00016} {\path{arXiv:2007.00016}}.
\newline\urlprefix\url{http://arxiv.org/abs/2007.00016}

\bibitem{Pathak2020-mz}
J.~Pathak, M.~Mustafa, K.~Kashinath, E.~Motheau, T.~Kurth, M.~Day,
  \href{http://arxiv.org/abs/2010.00072}{Using machine learning to augment
  {Coarse-Grid} computational fluid dynamics simulations} (Sep. 2020).
\newblock \href {http://arxiv.org/abs/2010.00072} {\path{arXiv:2010.00072}}.
\newline\urlprefix\url{http://arxiv.org/abs/2010.00072}

\bibitem{Frezat2020-zv}
H.~Frezat, G.~Balarac, J.~Le~Sommer, R.~Fablet, R.~Lguensat,
  \href{http://arxiv.org/abs/2010.04663}{Physical invariance in neural networks
  for subgrid-scale scalar flux modeling} (Oct. 2020).
\newblock \href {http://arxiv.org/abs/2010.04663} {\path{arXiv:2010.04663}}.
\newline\urlprefix\url{http://arxiv.org/abs/2010.04663}

\bibitem{Ling2016-qk}
J.~Ling, A.~Kurzawski, J.~Templeton,
  \href{https://www.cambridge.org/core/journals/journal-of-fluid-mechanics/article/reynolds-averaged-turbulence-modelling-using-deep-neural-networks-with-embedded-invariance/0B280EEE89C74A7BF651C422F8FBD1EB}{Reynolds
  averaged turbulence modelling using deep neural networks with embedded
  invariance}, J. Fluid Mech. 807 (2016) 155--166 (Nov. 2016).
\newline\urlprefix\url{https://www.cambridge.org/core/journals/journal-of-fluid-mechanics/article/reynolds-averaged-turbulence-modelling-using-deep-neural-networks-with-embedded-invariance/0B280EEE89C74A7BF651C422F8FBD1EB}

\bibitem{boris1973}
J.~P. Boris, D.~L. Book, Flux-corrected transport. i. shasta, a fluid transport
  algorithm that works, Journal of computational physics 11~(1) (1973) 38--69
  (1973).

\bibitem{book2012}
D.~L. Book, The conception, gestation, birth, and infancy of fct, in:
  Flux-Corrected Transport, Springer, 2012, pp. 1--21 (2012).

\bibitem{kingma2014}
D.~P. Kingma, J.~Ba, Adam: A method for stochastic optimization, arXiv preprint
  arXiv:1412.6980 (2014).

\bibitem{odman1997}
M.~T. Odman, A quantitative analysis of numerical diffusion introduced by
  advection algorithms in air quality models, Atmospheric Environment 31~(13)
  (1997) 1933--1940 (1997).

\bibitem{nair2010}
R.~D. Nair, P.~H. Lauritzen, A class of deformational flow test cases for
  linear transport problems on the sphere, Journal of Computational Physics
  229~(23) (2010) 8868--8887 (2010).

\bibitem{lauritzen2014}
P.~H. Lauritzen, A standard test case suite for two-dimensional linear
  transport on the sphere: results from a collection of state-of-the-art
  schemes, Geoscientific Model Development (2014).

\bibitem{saad2016}
T.~Saad, J.~C. Sutherland, Comment on “diffusion by a random velocity
  field”[phys. fluids 13, 22 (1970)], Physics of Fluids 28~(11) (2016) 22
  (2016).

\bibitem{zhuang2018}
J.~Zhuang, D.~J. Jacob, S.~D. Eastham, The importance of vertical resolution in
  the free troposphere for modeling intercontinental plumes, Atmospheric
  Chemistry and Physics 18~(8) (2018) 6039 (2018).

\bibitem{mcwilliams1984}
J.~C. McWilliams, The emergence of isolated coherent vortices in turbulent
  flow, Journal of Fluid Mechanics 146 (1984) 21--43 (1984).

\bibitem{methven1999}
J.~Methven, B.~Hoskins, The advection of high-resolution tracers by
  low-resolution winds, Journal of the atmospheric sciences 56~(18) (1999)
  3262--3285 (1999).

\bibitem{cheng2017}
Y.~Cheng, D.~Wang, P.~Zhou, T.~Zhang, A survey of model compression and
  acceleration for deep neural networks, arXiv preprint arXiv:1710.09282
  (2017).

\bibitem{carman2017}
J.~C. Carman, T.~Clune, F.~Giraldo, M.~Govett, A.~Kamrath, T.~Lee, D.~McCarren,
  J.~Michalakes, S.~Sandgathe, T.~Whitcomb, Position paper on high performance
  computing needs in earth system prediction, URL
  https://doi.org/10.7289/V5862DH3 (2017).

\bibitem{schulthess2018}
T.~C. Schulthess, P.~Bauer, N.~Wedi, O.~Fuhrer, T.~Hoefler, C.~Sch{\"a}r,
  Reflecting on the goal and baseline for exascale computing: a roadmap based
  on weather and climate simulations, Computing in Science \& Engineering
  21~(1) (2018) 30--41 (2018).

\bibitem{Note2}
Indeed, not all traditional solvers have ultra-low machine utilization. For
  example, the discontinuous Galerkin (DG) method can have a much higher
  machine utilization of 20\% \cite {breuer2019}. Although computational fluid
  dynamics (CFD) models can exploit high-order methods, climate models tend to
  use low-order schemes \cite {schulthess2018}.

\bibitem{Williams2009}
S.~Williams, A.~Waterman, D.~Patterson, Roofline: an insightful visual
  performance model for multicore architectures, Communications of the ACM
  52~(4) (2009) 65--76 (2009).

\bibitem{jouppi2017}
N.~P. Jouppi, C.~Young, N.~Patil, D.~Patterson, G.~Agrawal, R.~Bajwa, S.~Bates,
  S.~Bhatia, N.~Boden, A.~Borchers, et~al., In-datacenter performance analysis
  of a tensor processing unit, in: Proceedings of the 44th Annual International
  Symposium on Computer Architecture, 2017, pp. 1--12 (2017).

\bibitem{rasp2018}
S.~Rasp, M.~S. Pritchard, P.~Gentine, Deep learning to represent subgrid
  processes in climate models, Proceedings of the National Academy of Sciences
  115~(39) (2018) 9684--9689 (2018).

\bibitem{clima}
Clima: Climate machine, URL https://github.com/climate-machine/CLIMA.

\bibitem{breuer2019}
A.~Breuer, Y.~Cui, A.~Heinecke, Petaflop seismic simulations in the public
  cloud, in: International Conference on High Performance Computing, Springer,
  2019, pp. 167--185 (2019).

\end{thebibliography}

\end{document}